\documentclass{article}
\usepackage[utf8]{inputenc}
\usepackage{cite}
\usepackage{amsmath}
\usepackage{amssymb}
\usepackage{graphicx}
\usepackage{subcaption}
\usepackage{dsfont}
\usepackage{authblk}
\usepackage{hyperref}
\usepackage{siunitx}
\DeclareSIUnit\angstrom{\text{Å}}
\usepackage{xcolor}
\usepackage{caption}
\usepackage{textcomp}

\usepackage{float}
\usepackage{booktabs}
\usepackage{adjustbox}
\usepackage{tikz}
\usetikzlibrary{shapes, arrows, positioning,decorations.markings}
\tikzstyle{vecArrow} = [thick, decoration={markings,mark=at position
     1 with {\arrow[semithick]{open triangle 60}}},
double distance=1.4pt, shorten >= 5.5pt,
preaction = {decorate},
postaction = {draw,line width=1.4pt, white,shorten >= 4.5pt}]
\tikzstyle{innerWhite} = [semithick, white,line width=1.4pt, shorten >= 
4.5pt]
\hypersetup{hidelinks}
\usepackage[tmargin=1in,bmargin=1in,lmargin=0.75in,rmargin=0.75in]{geometry}
\newcommand{\R}{\ensuremath{\mathbb{R}}}
\newcommand{\N}{\ensuremath{\mathbb{N}}}

\providecommand{\keywords}[1]
{
     {
     \textbf{\textit{Keywords---}} #1
     }
}

\newcommand\blfootnote[1]{
     \begingroup
     \renewcommand\thefootnote{}\footnote{#1}
     \addtocounter{footnote}{-1}
     \endgroup
}

\title{Predicting the 3D microstructure of SOFC anodes from 2D SEM images using stochastic microstructure modeling and CNNs}
\author{Léon F. Schr\"oder$^{1,\ast}$, Sabrina Weber$^{1}$, Lukas Fuchs$^{1}$, Volker Schmidt$^{1}$, Benedikt Prifling$^{1}$\\}

\date{}

\begin{document}

\maketitle
\vspace{-4em}
\begin{center}
     \it
     $^1$Institute of Stochastics, Ulm University, 89069 Ulm, Germany
\end{center}
\blfootnote{$^\ast$Corresponding author \\ \textit{Email address: 
leon.schroeder@uni-ulm.de}}

\begin{abstract}
\noindent

The 3D microstructure of solid oxide fuel cell anodes significantly influences their electrochemical performance, but conventional methods for acquiring high-resolution  microstructural 3D data such as focused ion beam scanning electron microscopy (FIB-SEM) are costly in both time and resources. In contrast, obtaining 2D images, such as from scanning electron microscopy (SEM), is more accessible, though typically providing insufficient information to accurately characterize the 3D microstructure. To address this challenge, we propose a novel approach that predicts the 3D microstructure from 2D SEM images. The presented method utilizes a low-parametric 3D model from stochastic geometry to generate a large number of virtual 3D microstructures and employs a physics-based SEM simulation tool to obtain the corresponding 2D SEM images. By systematically varying the underlying model parameters, a large dataset can be generated to train convolutional neural networks (CNNs). By doing so, we can statistically reconstruct the 3D microstructure from 2D SEM images by drawing realizations from the stochastic 3D model using the predicted model parameters. In addition, we conducted an error analysis on key geometrical descriptors to quantitatively evaluate the accuracy and reliability of this stereological prediction tool.
\end{abstract}

\keywords{solid oxide fuel cell; anode; microstructure; FIB-SEM; stochastic 3D model;  SEM simulation; neural network}

\section{Introduction}

Environmentally friendly solid oxide fuel cells (SOFCs), a type of fuel cells that typically operates at temperatures between \SI{600}{\degreeCelsius} and \SI{1000}{\degreeCelsius}, are employed as a power source for buildings, industry and data centers due to their reliability and versatility \cite{brandon.2017, fergus.2016}. They are among the most energy-efficient sources, with an electrical efficiency of 60\% \cite{fergus.2016}, which can be increased to 90\% through the utilization of waste heat for heating purposes. 
One promising ionically conducting material in SOFC anodes is cerium-gadolinium oxide (CGO), often also known as gadolinium-doped ceria (GDC) \cite{bischof2019microstructure, sciazko2019influence, sciazko2019evaluation, hussain.2020}, which is used in combination with nickel (Ni) that is responsible for electron transport. 

\smallskip

Compared to the state-of-the-art Ni-YSZ (yttria-stabilized zirconia) anode, the Ni-CGO anode exhibits a higher oxide ion conductivity, especially at lower temperatures. Furthermore, in contrast to YSZ, CGO is a mixed ionic-electronic conductor (MIEC), which means that electronic and ionic conduction occur simultaneously in the anode \cite{eguchi1992electrical}. This leads to enhanced fuel utilization by enabling a much larger reaction zone compared to the the Ni-YSZ anode, where reactions only take place at the triple phase boundary (TPB), i.e. the interface between pore space, nickel and YSZ. Other advantages are the higher resistance to carbon deposition \cite{kubota2017self} and sulfur poisoning \cite{zhang2010comparative}.

\smallskip

In addition to material composition, the 3D microstructure of the anode also has a major impact on its electrochemical performance \cite{lee2003impact}. It can be altered to achieve, among others,  higher efficiency and less pronounced degradation behavior, which is essential to maximize the potential of SOFCs as a green energy solution. A suitable imaging tool to investigate the microstructure is scanning electron microscopy (SEM). However, SEM provides only 2D images, which limits its ability to capture the geometrically complex 3D microstructure. For a full 3D reconstruction, focused ion beam scanning electron microscopy can be performed. This technique sequentially mills the specimen layer by layer with a focused ion beam (FIB), capturing a series of 2D SEM images that can be compiled into a 3D image \cite{banhart.2008}. However, this approach is destructive and yields rather limited data as a result of its high costs in time and resources.

\smallskip

In contrast to 3D FIB-SEM tomography, 2D SEM images can be acquired quickly at a low cost, which motivates the investigation of methods for predicting 3D microstructures based solely on 2D SEM images. However, a large amount of training data will be needed to train a reliable predictor that uses a 2D SEM image as input to predict the corresponding 3D microstructure. Thus, to obtain a sufficiently large amount of data, we generate virtual microstructures using a stochastic 3D modeling approach. In particular, a low number of parameters serves as an interpretable representation of the 3D microstructure. Moreover, through systematic variation of the stochastic model's parameters, an extensive amount of virtual but realistic 3D microstructures can be generated.
These microstructures can then be used as 3D geometry input for Nebula \cite{van2020nebula}, a 2D SEM simulation tool that simulates electron trajectories by physically simulating electron-matter interactions. This approach yields a dataset comprising 2D SEM images and their respective 3D model parameters, which is used as ground truth for training convolutional neural networks (CNNs) that predict the 3D model parameters based on a single 2D SEM image. 

\smallskip

More precisely, this paper explores how CNNs, a class of deep learning models, can be applied to estimate the parameters of the stochastic 3D model and, consequently, predict the 3D microstructure of SOFC anodes based on 2D SEM images. Note that three CNNs will be trained separately, where each CNN will predict a certain predefined subset of the parameter vector of the stochastic 3D microstructure model, see Section~\ref{sec:netarch} for further details. CNNs are commonly designed to process and analyze images, e.g. for image recognition and image classification tasks. By applying layers of convolutional filters, CNNs can automatically learn spatial hierarchies of features, from simple edges and textures to complex shapes and patterns. During training, the weights between the layers are adaptively tuned to detect these spatial features, effectively building representations that allow accurate parameter estimation. There are several network architectures well known for their effectiveness, e.g. VGG \cite{vgg_simonyan}, MobileNet \cite{mobilenetv3} and ResNet \cite{he2016deep}. While VGG is an easy-to-understand and deep network, MobileNet reduces the computational cost and the number of parameters by using depthwise separable convolutions, and ResNet addresses the vanishing gradient problem with its use of residual connections.

\smallskip

The approach considered in the present paper offers significant practical advantages for academia and industry, allowing substantial savings in both time and cost by eliminating the need for 3D tomography. 
Furthermore, the CNN-based prediction of the parameters of the stochastic 3D microstructure model, where the parameters are predicted by three different CNNs, also avoids the phase-based segmentation (into pore space, nickel and electrolyte materials such as YSZ or CGO) of 2D SEM images.
Thus, this framework provides a promising tool for accurately characterizing the 3D microstructure of SOFC anodes, offering significant savings in time and cost compared to traditional tomographic imaging.

\smallskip

\begin{figure}[H]
    \centering
    \includegraphics[width=0.63\linewidth]{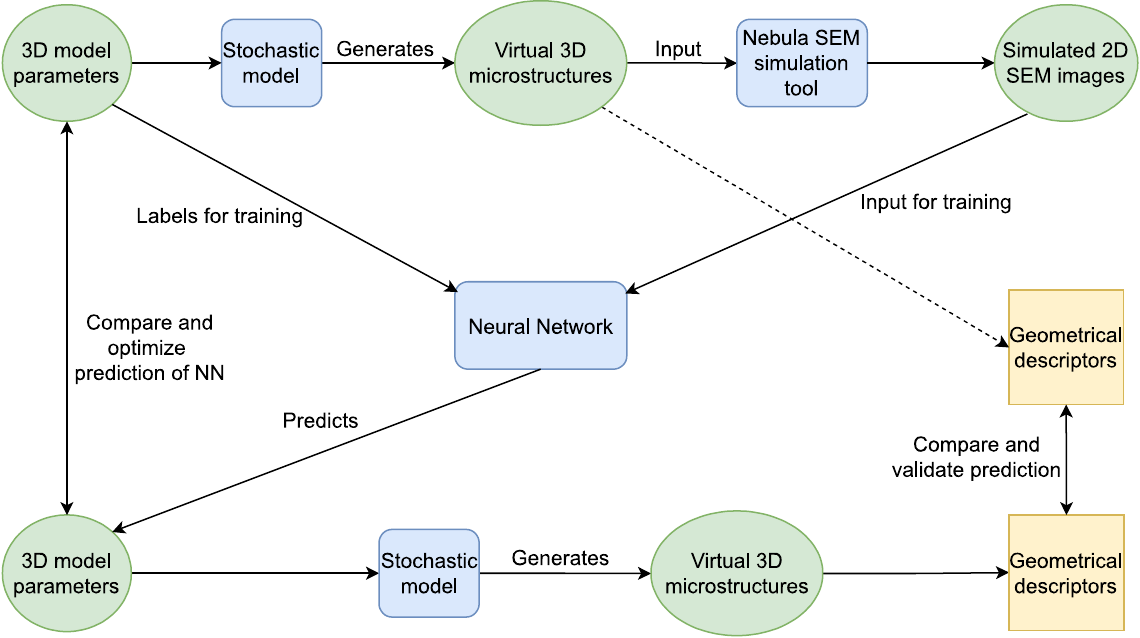}
    \caption{\textbf{Workflow of the present paper.}}
    \label{fig:workflow}
\end{figure}

The performance of this approach is evaluated by comparing geometrical descriptors of 3D microstructures drawn from the stochastic 3D model with preset parameters (treated as ground-truth information) with geometrical descriptors of virtual 3D microstructures that have been generated by the stochastic model using the parameters predicted by the CNNs. The whole workflow including image generation, training, and validation is visualized in Figure~\ref{fig:workflow}.

The rest of this paper is structured as follows. In Section~\ref{sec:microstructure_modeling}, the stochastic model that is used to generate virtual 3D microstructures is presented. Then, in Section~\ref{sec:nebula}, we describe the application of the Nebula SEM simulation tool that is used to create virtual 2D SEM images. Section~\ref{sec:cnn} explains the neural networks' architecture, training procedure, and data augmentation techniques. The results obtained using these neural networks are then analyzed in Section~\ref{sec:results}, where the performance of the prediction tool is evaluated with respect to geometrical descriptors. A summary of the main results and an outlook for possible future research activities is presented in Section~\ref{sec:conclusion}.

\section{Stochastic 3D microstructure modeling}
\label{sec:microstructure_modeling}

In this section, we introduce several geometrical descriptors of 3D microstructures considered in the present paper, state a parametric stochastic 3D microstructure model, and explain the generation of virtual 3D microstructures by varying the parameters of the stochastic model.

\subsection{Geometrical descriptors}\label{sec:descriptors}

Table~\ref{tab:descriptors} contains an overview of five geometrical descriptors that will be used to quantitatively evaluate the prediction accuracy of the presented framework, namely the volume fraction $\varepsilon_i$, the specific surface area (SSA) $S_i$, the mean chord length $\mu_i$, the mean geodesic tortuosity $\tau_i$ of the $i$-th phase for $i\in\{\text{Ni},\text{P},\text{CGO}\}$, and the specific length of the TPB. A detailed explanation of these geometrical descriptors can be found in \cite{weber_descriptors}.

\begin{table}[H]
    \centering
    \begin{tabular}{llll}
    \toprule
     Descriptor & Symbol & Unit & Range of possible values\\ 
     \midrule
     Volume fraction & $\varepsilon_i$ & [ ] & $[0,1]$\\
     Specific surface area & $S_i$  & [\si{\per\micro\meter}] & $[0,\infty)$\\
     Mean chord length & $\mu_i$ & [\si{\micro\meter}] & $[0,\infty)$ \\
     Mean geodesic tortuosity & $\tau_i$ & [ ] & $[1,\infty)$ \\
     Specific length of TPB & $\rho$ & [\si{\micro\meter^{-2}} ] & $[0,\infty)$ \\
     \bottomrule
    \end{tabular}
    \caption{\textbf{Overview of the geometrical descriptors considered in the present paper.} Note that the index $i$ will be later replaced by the abbreviation of the phase name, for which the geometrical descriptor is computed.}
    \label{tab:descriptors}
\end{table}

\subsection{Model description}
\label{sec:model_description}

We now describe the stochastic model that is used to generate virtual 3D microstructures of SOFC anodes.
The main building block of the model consists of two stochastically independent and motion-invariant Gaussian random fields $X = \{ X(t), t \in \mathbb{R}^3 \}$ and $Y = \{ Y(t), t \in \mathbb{R}^3 \}$, with the covariance functions $\rho_X,\rho_Y: [0,\infty) \rightarrow \mathbb{R}$, given by $\rho_X(h) = \text{Cov}(X(s), X(t))$ and $\rho_Y(h) = \text{Cov}(Y(s), Y(t))$ for each $h>0$, where $s,t \in \R^3$ with $|s-t| = h$ \cite{chiu2013stochastic}. Note that for motion-invariant random fields, the covariance functions only depend on the distance $h$ of the two points $s$ and $t$. Furthermore, we assume that $X$ and $Y$ are standardized, i.e., we assume that $\mathbb{E}X(t) = \mathbb{E}Y(t) = 0$ and Var$X(t)$ = Var$Y(t)$ = 1 holds for all $t \in \mathbb{R}^3$.
Next, we consider the $\chi^2$-field with two degrees of freedom, denoted by $Z = \{Z(t), t \in \mathbb{R}^3 \}$, which is defined by
$Z(t) = X_1^2(t) + X_2^2(t)$ for each $t\in \R^{3}$, where $X_1, X_2$ are stochastically independent copies of $X$. 
Using two threshold parameters, $\lambda_Y, \lambda_Z \in \mathbb{R}$, we then define the following three random closed sets as:
\begin{equation}
\begin{aligned}
    \Xi_{\text{Ni}} &= \{t\in \mathbb{R}^3: Z(t) \geq \lambda_Z \} \\
    \Xi_{\text{P}} &= \text{cl}(\{t\in \mathbb{R}^3: Y(t) \geq \lambda_Y \} \cap \Xi_1^\complement)\\
    \Xi_{\text{CGO}} &= \text{cl}((\Xi_{\text{Ni}} \cup \Xi_{\text{P}})^\complement),
\end{aligned}\label{eq:one}
\end{equation}
where $\text{cl}$ and $\complement$ denote the topological closure and the complement of a set, respectively. More information on the theory of random closed sets can be found, for example, in \cite{molchanov2005theory}. Note that the random closed sets $\Xi_{\text{Ni}}$, $\Xi_{\text{P}}$ and $\Xi_{\text{CGO}}$ stated in Eqs.~\eqref{eq:one} model the nickel phase, the pore space and the CGO phase, respectively. To obtain a fully parametric microstructure model, we assume that the covariance functions $\rho_{X}$ and $\rho_{Y}$ of $X$ and $Y$ belong to the exponential family, that is, we have $\rho_X(h)=\exp(-(p_1 h)^{p_2})$ and $\rho_Z(h)=\exp(-(p_3 h)^{p_4})$ for each $h\geq 0$ with model parameters $p_1, p_2, p_3, p_4 >0$.
\cite{lantuejoul.2013}. In total, the final model is characterized by six parameters: $\lambda_Y, \lambda_Z, p_1, p_2, p_3,p_4$. By systematically varying these parameters, we are able to obtain a large set of diverse virtual 3D microstructures, which is described in the following section.

\subsection{Generation of virtual 3D microstuctures}
\label{sec:simulation_study}

Virtual 3D microstructures of size $400 \times 400 \times 400$ voxels with a voxel size of \SI{50}{\nano\meter} have been drawn from the stochastic model described in Section ~\ref{sec:model_description}. To create a diverse database of virtual but realistic SOFC anode structures, the parameters $p_1$ and $p_3$ were chosen uniformly from the interval $[0.05, 0.15]$, while $p_2$ and $p_4$ were chosen uniformly from the interval $[1.6, 2]$. To obtain reasonable microstructures, candidates of the three volume fractions $\varepsilon_1,\varepsilon_2,\varepsilon_3$ were drawn uniformly from $[0.1,0.8]$ and subsequently normalized by their sum to ensure that $\varepsilon_1+\varepsilon_2+\varepsilon_3=1$. Any realizations where at least one volume fraction fell below $0.1$ were systematically discarded. Then, using the sampled volume fractions of the three phases, the parameters $\lambda_Y$ and $\lambda_Z$ were determined via 
\begin{equation}
    \lambda_{Z} = \chi^{-1}(1-\varepsilon_{\text{Ni}}) \text{ and } \lambda_{Y} = \Phi^{-1}(1 - \frac{\varepsilon_{\text{P}}}{1-\varepsilon_{\text{Ni}}}),
\end{equation}
where $\chi^{-1}:(0,1) \to \R$ and $\Phi^{-1}:(0,1) \to \R$ denote the quantile function of a $\chi^{2}$ distribution with two degrees of freedom and a standard normal distribution, respectively. These six model parameters define the ground truth labels that are used to train the neural networks. In total, 2000~microstructures have been generated to ensure sufficient structural diversity.

\section{Physics-based simulation of 2D SEM images}
\label{sec:nebula}

After generating a wide range of 3D microstructures as stated in Section~\ref{sec:microstructure_modeling}, we now describe the simulation of the 2D SEM images that use these 3D structures as input. One of the most recent and physically accurate SEM simulation tools is the Nebula simulator proposed in \cite{van2020nebula}, which will be used in this paper.

\subsection{SEM simulation setup}\label{subsec:SEM}

3D voxelized SOFC anode structures can be passed to Nebula by meshing them into a triangle format, where the interface area between two phases is represented as a union of triangles along with the information on the materials that make up this interface~\cite{van2020nebula}. This meshing is performed by cutting the interface area between two voxels that belong to different phases diagonally into two triangles. In addition to the SOFC anode structure, certain unique materials are included in the triangle file. First, there is the backscattered electron detector, which is in our case located five centimeters above the sample. When an electron reaches the detector, the simulation of this path ends and it is recorded into an output file if its energy is above the threshold of \SI{50}{\eV}. A perfect mirror is an additional special material that is affixed to the sample's outer boundary points. The electrons are maintained in space by these mirrors, enabling the simulator to terminate.
When an electron penetrates the whole sample, a terminator placed below the sample ends the electron's route. Also, this is a computational necessity since, in the absence of it, electrons might never reach a terminal state.

\smallskip

Furthermore, a material file needs to be created for each material used in our example, nickel and CGO. For the pore space, this step is not necessary, as it is assumed to be a vacuum by default, which is already included in the software. By providing a parameter file to the scattering cross section tool (cstool), which depends on the ELSEPA software package~\cite{salvat2005elsepa}, this can be accomplished. These parameters are used by cstool to characterize the electron scattering behavior in this material, using the physics models described in \cite{kieft2008refinement} and \cite{penn1987electron}. 
The parameter file contains various physical characteristics, where the most important parameters for the three materials are depicted in Table~\ref{tab:material_parameters}.

\begin{table}[H]
    \centering
    \begin{tabular}{lll}
    \toprule
     Material parameter & Nickel & CGO \\ \midrule
     Atomic number [\ ] from \cite{greenwood.1998} & 28 & 58, 64, 8 \\
     Atomic mass [$\mathsf{Da}$] from \cite{greenwood.1998} & 58.7  & 140.1, 157.3, 16 \\
     Density [$\SI{}{\gram\per\centi\meter^3}$] & 8.9 \cite{greenwood.1998} & 7.2 \cite{mishra.2019} \\
     Lattice parameter [$\SI{}{\angstrom}$] & 3.5 \cite{arblaster.2018} & 5.45 \cite{sistla.2019} \\
     \bottomrule
    \end{tabular}
    \caption{Material parameters for nickel and CGO ($\mathrm{Ce}_{0.8}\mathrm{Gd}_{0.2}\mathrm{O}_{2}$). The atomic mass is given in Dalton, where $1\ \mathsf{Da} \approx \SI{1.66e-27}{\kilo\gram}$. The atomic number and mass are given for all elements contained in the material, i.e., Ce, Gd and O for CGO (from left to right).}
    \label{tab:material_parameters}
\end{table}

In addition to this, the electrons whose paths we want to simulate must be defined. The grid points, i.e. the points where the electrons initially are shot at, are selected as the voxel center points on the structure's surface.
The image becomes sharper, and the amount of noise in the data decreases as we use more electrons. Since authentic SEM images contain some noise, this is beneficial up to a certain extent. Also, note that the computational cost increases linearly with the quantity of electrons. With this parameter, we can match the noise of the simulated images with the noise of tomographic image data of a real SOFC anode obtained by 3D FIB-SEM. More precisely, we consider image data from sample A in \cite{weber_descriptors} as reference for determining the number of electrons. To achieve this, we take an original greyscale 3D FIB-SEM image $\tilde{A_1}: [0,w_1] \times [0,w_2] \times [0,w_3] \cap \mathbb{N}_0^3 \rightarrow \N_0$ with $w_{1}, w_{2}, w_{3} \in \N$, from which we extract the top slice $I_1: [0,w_1] \times [0,w_2] \cap \mathbb{N}_0^2 \rightarrow \mathbb{N}_0$,  where $\mathbb{N}=\{1,2,\ldots\}$ and $\mathbb{N}_0=\{0\}\cup\mathbb{N}$.
From $\tilde{A_1}$, the corresponding segmented image $A_1: [0,w_1] \times [0,w_2] \times [0,w_3] \cap \mathbb{N}_0^3 \rightarrow \{0,127,255\}$ has also been provided, where the values 0, 127 and 255 correspond to pore space, nickel, and CGO, respectively. We then simulate a grayscale SEM image $I_2: [0,w_1] \times [0,w_2] \cap \mathbb{N}_0^2 \rightarrow \mathbb{N}_0$ of the structure $A_1$ via Nebula. Next, let $C=\{(x,y) \in [0,w_1] \times [0,w_2] \cap \mathbb{N}_0^2: A_1(x,y,w_3) = 255\}$ be the set of all pixels that have been classified as the CGO phase in the top layer of the original structure. To quantify the noise within the CGO phase, we consider the sample mean
\begin{equation}
    \bar{x}(I) = \frac{\sum_{(x,y) \in C}I(x,y)}{|C|}
\end{equation}
and sample standard deviation 
\begin{equation}
    s(I) = \sqrt{\frac{\sum_{(x,y) \in C} (I(x,y) - \bar{x}(I))^2}{|C|-1}}
\end{equation}
of an image $I: [0,w_1] \times [0,w_2] \cap \mathbb{N}_0^2 \rightarrow \mathbb{N}_0$ on the set $C$ to estimate the coefficient of variation 
\begin{equation}
    \widehat{cv}(I) = \frac{s(I)}{\bar{x}(I)},
\end{equation}
for $\bar{x}(I) \neq 0$, where $|C|$ denotes the cardinality of $C$.
We computed this measure for the original SEM image $I_1$ and performed a binary search by varying the number of electrons used to simulate grayscale SEM images $I_2$ until both estimated coefficients of variation match.
As a result, a total of 1400 electrons will be virtually fired onto the sample for each grid point.

\smallskip

Since the electrons move in a preset, straight path downward in a negative z-direction towards the sample and do not lose energy in the vacuum until they can interact with the material, the distance between them and the sample is of little significance. Moreover, an acceleration voltage of \SI{5000}{\volt} is used such that electrons can penetrate into the sample, leading to shine-through artifacts. Secondary electrons may form during the scattering. They will be handled in the same manner as the primary electrons and added to the list of active electrons.\\

\begin{figure}[H]
    \centering
    \raisebox{-4mm}{
    \begin{subfigure}[t]{0.27\textwidth}
        \centering
        \includegraphics[width=\textwidth]{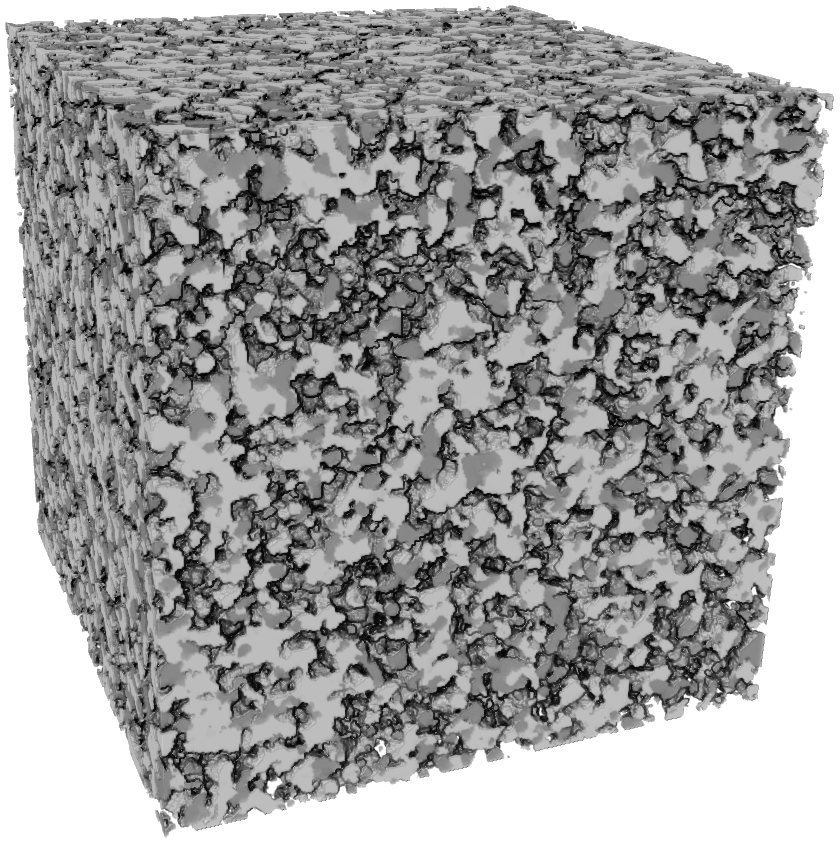}
    \end{subfigure}
    }
    \hspace{1cm}
    \begin{subfigure}[t]{0.25\textwidth}
        \centering
        \includegraphics[width=\textwidth]{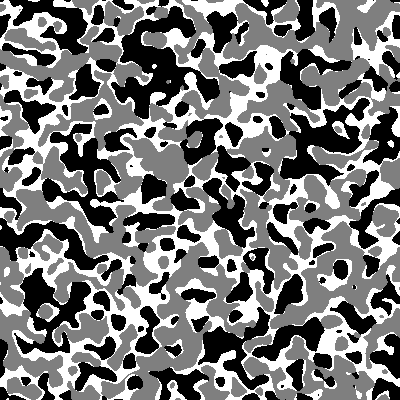}
    \end{subfigure}
    \hspace{1cm}
    \begin{subfigure}[t]{0.25\textwidth}
        \centering
        \includegraphics[width=\textwidth]{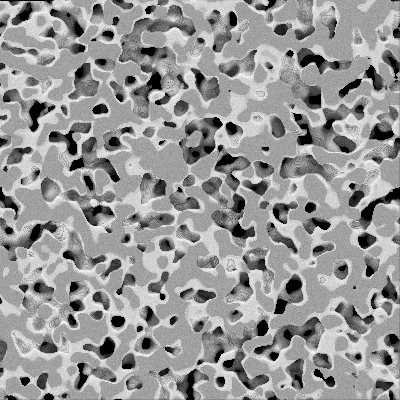}
    \end{subfigure}
    \caption{\textbf{Stochastic microstructure model realization and corresponding SEM image.} Realization of the stochastic 3D microstructure model (left) of size $\SI{20}{\micro\meter} \times \SI{20}{\micro\meter} \times \SI{20}{\micro\meter}$, top slice ($\SI{20}{\micro\meter} \times \SI{20}{\micro\meter}$) of the realization (middle) and the corresponding SEM image ($\SI{20}{\micro\meter} \times \SI{20}{\micro\meter}$) obtained by Nebula using the settings described in Section~\ref{subsec:SEM}. In the 3D rendering, the pore space is transparent, the CGO phase is depicted in the brighter gray and nickel is colored in dark gray, whereas in the 2D slice, the pore space is black, the CGO phase is white and the nickel phase is gray.}
    \label{fig:nebula_sem}
\end{figure}

\subsection{Simulation of virtual 2D SEM images}
Using the database of 2,000 generated microstructures (see Section~\ref{sec:simulation_study}), we applied the SEM simulation to each structure, an example of which is shown in Figure~\ref{fig:nebula_sem}. The simulation of a single SEM image for a $400 \times 400 \times 400$ structure with $1400$ electrons per pixel with an energy of \SI{5000}{\eV}, that is, a total of $1400 \cdot 400 \cdot 400 =  224 \cdot 10^6$ electrons, takes roughly five minutes on an NVIDIA GeForce RTX 4060 Ti. From this process, we obtained the dataset $G = \{(A_i, I_i, \theta_i): 1 \leq i \leq 2000\}$, where $A_i: [0,400]\times[0,400]\times[0,400] \cap \mathbb{N}_0^3 \rightarrow \{0,127,255\}$ is the virtual 3D microstructure, $I_i: [0,400]\times[0,400] \cap \mathbb{N}_0^2 \rightarrow \N_0$ denotes the simulated SEM image of $A_i$ and $\theta_i = (\lambda_Y, \lambda_Z,p_1, p_2, p_3, p_4)_i$ contains the model parameters that have been used to generate $A_i$.
The set $G$ was then divided into a training set $T$ with 1600 microstructures (80\%) and a validation set $V$ containing the remaining 400 microstructures (20\%).

\section{Convolutional neural networks for prediction of model parameters}
\label{sec:cnn}
In this section, the neural networks that predict the model parameters of the stochastic 3D microstructure model from 2D SEM images are described. First, a CNN extracts features from the input SEM images, which are then processed by a subsequent neural network that estimates the 3D model parameters. We provide a detailed overview of the network architecture, including its layer composition, activation functions, and parameter grouping strategy. The networks are trained supervised, meaning that they are trained such that their outputs predict the ground truth labels, i.e., the 3D model parameters. For supervised training, we exclusively used the training set $T$ and evaluated its performance on the unseen validation set $V$ to assess its generalization performance; see Section~\ref{sec:results}.

\subsection{Data augmentation}\label{sec:data_augmentation}

To increase the diversity of the training dataset and improve the robustness of the parameter prediction model, we apply data augmentation techniques to the 2D SEM images that are used as input for the neural networks. Specifically, we first randomly rotate them around the center of the image, by uniformly sampling angles from $\{0, \frac{\pi}{2}, \pi, \frac{3\pi}{2}\}$. Moreover, from an image $I$ of size $400 \times 400$, we cut five pixels at the boundaries to mitigate possible boundary effects that could interfere with training and randomly select a cutout of size $256 \times 256$ by sampling starting points $x_s$ and $y_s$ uniformly from $\{5, \ldots , 138\}$, which  yields the cutout $I_{\rm cut}:[x_s,x_s+255]\times [y_s,y_s+255]\cap\N_0^2\to\N_0$. 

\smallskip

Although the pixel intensities in the simulated SEM image directly correspond to the number of detected electrons, real SEM images may exhibit differing pixel intensity distributions due to various post-processing steps used by the software of the FIB-SEM device. To ensure comparability between these real SEM images and our simulated SEM images and to improve the generalizability and accuracy of the model to unseen data \cite{sinsomboonthong2022performance}, the values of the SEM images are first normalized to the range of $[-0.5, 0.5]$ using min-max normalization, i.e., $I_{\text{norm}} = \frac{I-\text{min}(I)}{\text{max}(I)-\text{min}(I)} - 0.5$. The values are then rounded to 256 equidistant values.

\begin{figure}[H]
    \centering
    \includegraphics[width=0.6\linewidth]{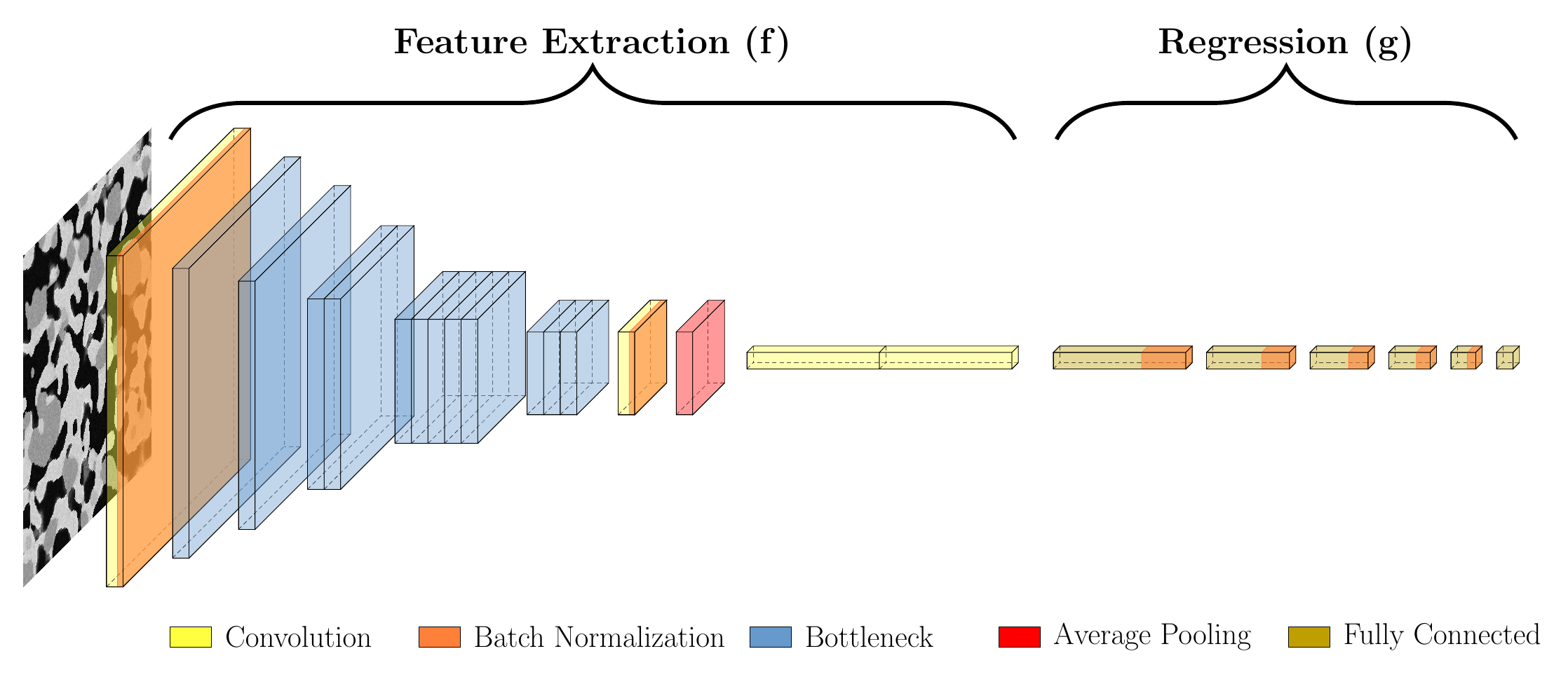}
    \caption{\textbf{Schematic visualization of the network architecture. }The subnetwork $f$ extracts features out of the input image, whereas subnetwork $g$ processes these features to predict the model parameters.}
    \label{fig:schematic_cnn}
\end{figure}

\subsection{Network architecture and training}\label{sec:netarch}

The architecture of our neural networks $\text{CNN}_i: \mathbb{R}^{256 \times 256} \rightarrow \mathbb{R}^{2}$, for $i \in \{1,2,3\}$, consists of two concatenated subnetworks, denoted as \( f_i: \mathbb{R}^{256 \times 256} \rightarrow \mathbb{R}^{1000} \) and \( g_i: \mathbb{R}^{1000} \rightarrow \R^2 \), i.e., $\text{CNN}_i = g_i(f_i)$. The subnetwork \( f_i \) takes the SEM images, which have been processed as described in Section~\ref{sec:data_augmentation}, as input and serves as a feature extraction network. These extracted features are used as input for \( g_i \), which acts as a regression network to predict the parameters of the 3D stochastic model. A schematic visualization of the network architecture, illustrating the two-stage process of feature extraction and parameter prediction, is shown in Figure~\ref{fig:schematic_cnn}.

\smallskip

For the feature extraction network $f_i$, we use as a pretrained backbone network, that is, a deep neural network that has already been trained on a large dataset. This has the advantage that it significantly reduces training time and computational costs. We chose a version of MobileNetV3 as our backbone network, which consists of stacked convolutional \cite{lecun1998gradient}, bottleneck \cite{he2016deep}, squeeze-and-excitation \cite{hu2018squeeze}, batch normalization \cite{ioffe2015batch} and pooling layers. By using computationally efficient activation functions and layer structures, it significantly reduces the number of parameters and thus is computationally fast. Note that the modular structure allows an easy replacement of the feature extraction network $f_i$ by a different architecture. For a detailed overview of the chosen architecture, the reader is referred to \cite{mobilenetv3}. Overall, it takes a $256 \times 256$ cutout of the SEM image as input and maps it to a feature vector of size $1000$.

\smallskip

This output of $f_i$ is then processed further as it serves as input to the regression network $g_i$. The network $g_i$ is composed of dense (fully connected) layers that linearly map an input vector $X \in \R^k$ to an output vector $Y \in \R^\ell$ using a weight matrix $W \in \R^{\ell \times k}$ for some $k,\ell\in\mathbb{N}$, i.e., $Y = W  X$. A dense layer is followed by a batch normalization and subsequently passed through the ReLU (rectified linear unit) activation function to introduce non-linearity, where the ReLU function $\text{ReLU}: \R \to [0,\infty)$ is defined as 
\begin{equation}
   \text{ReLU}(x) = \max\{0,x\} \text{ for each } x \in \R.
\end{equation}
For the last layer the use of batch normalization and activation functions is omitted to not constrain the range of the output. Overall, $g_i$ takes the features extracted by $f_i$ and outputs a subset of the 3D model parameters $(\lambda_Y, \lambda_Z, p_1, p_2, p_3, p_4)$. 

\smallskip

Since the parameters of the stochastic 3D model are defined on different scales ($\lambda_{Y}, \lambda_{Z} \in \R, p_{1},p_{3} \in [0.05,0.15], p_{2},p_{4} \in [1.6,2])$, the model parameters are divided into three groups based on their scale. The first group, which contains $\lambda_Y$ and $ \lambda_Z$, is predicted by the neural network $\text{CNN}_1$. A second network, $\text{CNN}_2$, predicts the covariance parameters $p_1$ and $p_3$, while a third neural network $\text{CNN}_3$ predicts the covariance parameters $p_2$ and $p_4$. This grouping strategy mitigates the risk of parameter imbalance, where parameters with larger magnitudes could otherwise dominate the learning process, leading to suboptimal predictions for those with smaller scales. The network $\text{CNN}: \R^{256 \times 256} \rightarrow \R^6$ that predicts all six parameters is then defined by combining the three networks, i.e., $\text{CNN} = (\text{CNN}_1, \text{CNN}_2, \text{CNN}_3)$.

\smallskip

For an input $x = (x_1, \ldots, x_B)\in\R^B$ of some batch size \( B \in \mathbb{N} \), the networks predict an output $\hat{y} = (\hat{y}_1, \ldots, \hat{y}_B) \in\R^B$, where \( \hat{y}_i = \text{CNN}(x_i) \) for each \( i \in \{1, \ldots, B\} \). The prediction \( \hat{y} \) is compared with the ground truth $ y = (y_1, \ldots, y_B) \in\R^B$ using the mean squared error (MSE) as a loss function:
\[
\text{MSE}(y, \hat{y}) = \frac{1}{B} \sum_{i=1}^B (y_i - \hat{y}_i)^2.
\]
This loss is minimized using the Adam optimizer \cite{kingma2017adammethodstochasticoptimization}, a gradient-based adaptive optimization algorithm, with a learning rate of 0.001. A batch size of $B = 128$ is used throughout the training.
\smallskip
 
Training is conducted for a maximum of 8000 epochs, with early stopping applied to prevent overfitting. Specifically, the training halts if there is no improvement in the training loss for 500 consecutive epochs, effectively reducing the risk of overfitting, improving generalization and saving training time.

\section{Results}
\label{sec:results}

In this section, we present the results of the model parameter prediction. On the one hand, we directly investigate the goodness of fit with respect to the parameters of the stochastic 3D microstructure model. On the other hand, we analyze the accuracy of the geometric descriptors computed from 3D structures that have been simulated using the parameters predicted by the CNNs. To ensure the validity and generalizability of the neural network, all results in the following two subsections were performed exclusively on the validation set $V$. Finally, the accuracy with respect to selected geometrical descriptors is also computed for tomographic image data of a SOFC anode to investigate how well the prediction tool performs on real-world data.

\subsection{Evaluation with respect to model parameters}
\label{subsec:eval_params}

To measure the accuracy of the prediction, we both consider the MSE as well as the mean absolute error (MAE), which is the mean absolute deviation of the ground truth $y = (y_1, \ldots, y_B)\in\R^B$ and the prediction $\hat{y} = (\hat{y}_1, \ldots, \hat{y}_B)\in\R^B$ for some batch size $B \in \N$, mathematically defined as

\begin{equation*}
    \text{MAE}(y, \hat{y}) = \frac{1}{B} \sum_{i=1}^B |y_i - \hat{y}_i|.
\end{equation*}

\begin{figure}[H]
    \centering
    \begin{subfigure}{0.32\textwidth}
        \centering
        \includegraphics[width=\linewidth]{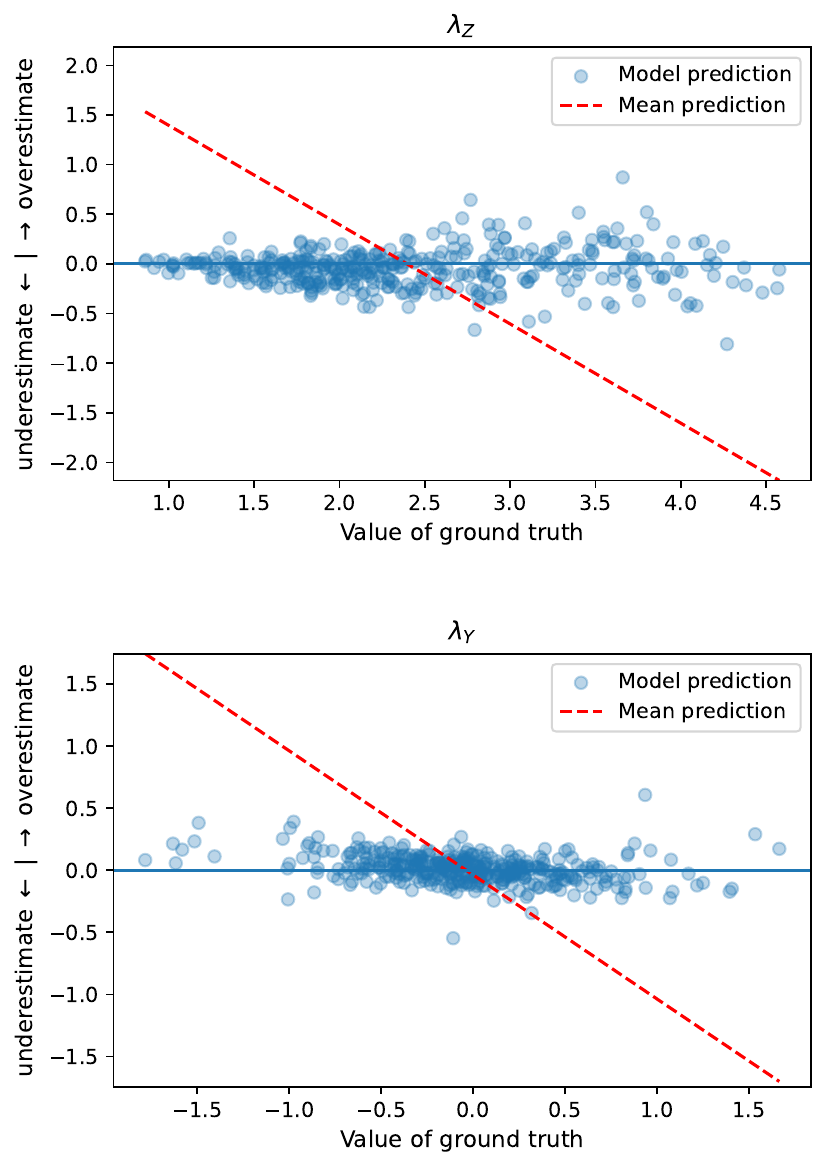}
        \label{fig:err_lambdas}
    \end{subfigure}
    \begin{subfigure}{0.32\textwidth}
        \centering
        \includegraphics[width=\linewidth]{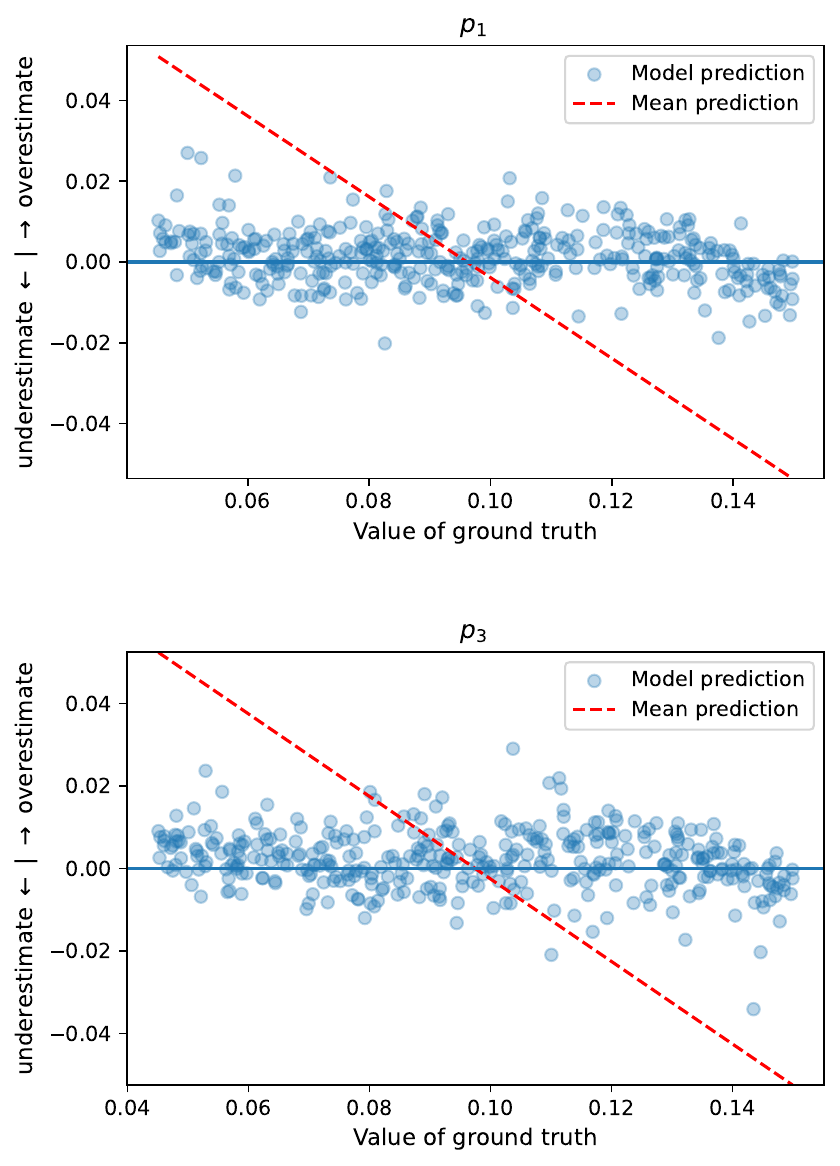}
        \label{fig:err_covs1}
    \end{subfigure}
    \begin{subfigure}{0.32\textwidth}
        \centering
        \includegraphics[width=\linewidth]{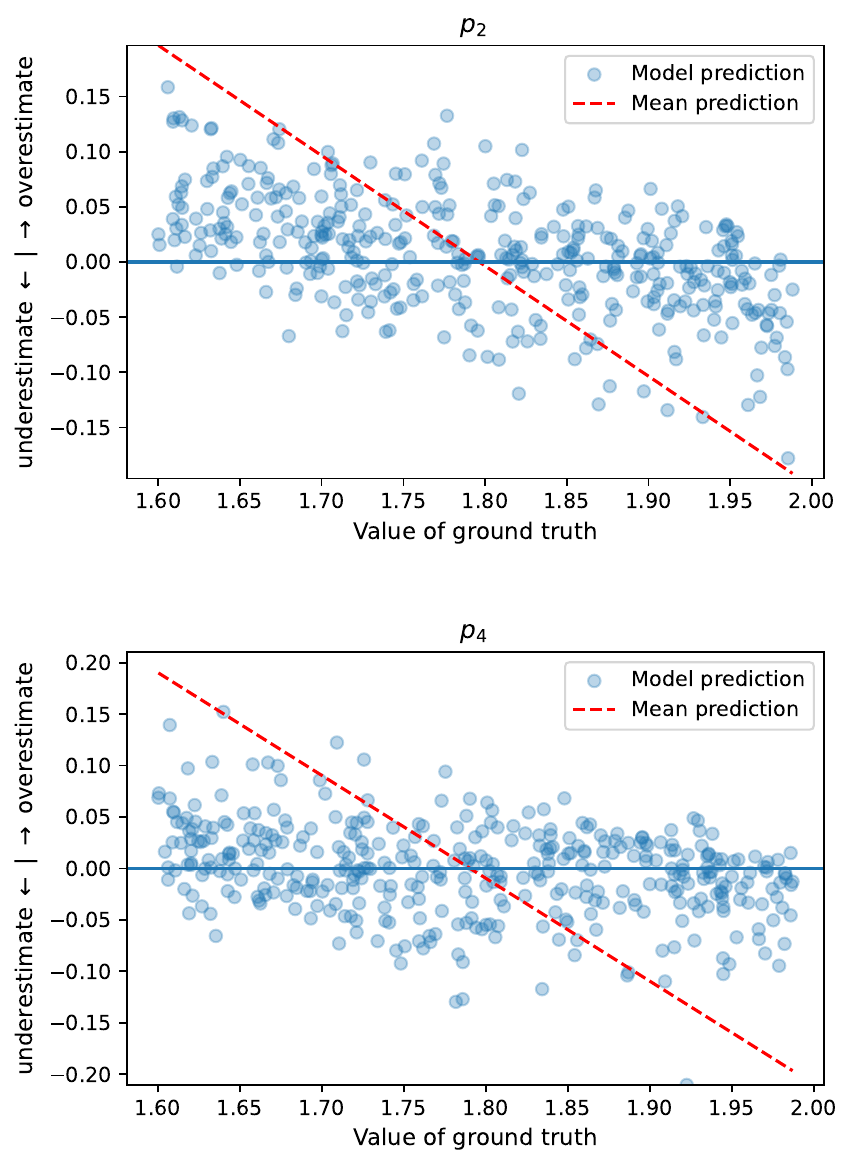}
        \label{fig:err_covs2}
    \end{subfigure}
    \caption{\textbf{Accuracy of CNN predictions.} Comparison of ground truth and difference between prediction and ground truth for the parameters $\lambda_Z, \lambda_Y$ (left), $p_1, p_3$ (middle) and $p_2, p_4$ (right) on the validation set $V$. Positive values on the y-axis indicate an overestimation of the model parameter by the CNNs, while negative values show an underestimation.}
    \label{fig:combined_errors_params}
\end{figure}

Figure~\ref{fig:combined_errors_params} shows the deviation of the predicted parameters from the ground truth parameters. Note that the red dashed line corresponds to the error that would result from a constant prediction, where for each input the arithmetic mean of all training labels is predicted. This approach will be called mean prediction in the following and will serve as a baseline for assessing the predictive power. 

\smallskip

With an MAE of 0.1345 and an MSE of 0.0332 for $\lambda_Y$ and an MAE of 0.0804 and an MSE of 0.0106 for $\lambda_Z$, the two threshold parameters provide a good fit. We can observe a significant increase of accuracy for the CNN prediction compared to the mean prediction. 
Similarly, the covariance parameters $p_1$ (MAE: 0.00516, MSE: $4.202 \cdot 10^{-5}$) and $p_3$ (MAE: 0.00548, MSE: $4.859 \cdot 10^{-5}$) can be well predicted. 
In contrast, predicting $p_2$ (MAE: 0.0394, MSE: 0.00247) and $p_4$ (MAE: 0.0346, MSE: 0.002) turned out to be more challenging.

Overall, the CNNs are able to accurately predict the parameters of the stochastic 3D microstructure model, demonstrating the ability to generalize to unseen data. With respect to the application of this prediction tool, the effect of the parameters of the stochastic model on geometrical descriptors of the resulting 3D microstructure is even more important, and thus will be discussed in the following section. 

\subsection{Evaluation with respect to geometrical descriptors}
\label{subsec:eval_descr}
To ensure a robust prediction of the model parameters from a 2D SEM image $I_V \in V$, eight cutouts $I_V^1, \ldots, I_V^{8}$ of $I_V$ are selected at random. These cutouts are independently processed by the neural networks, and the final prediction of the model parameters is determined as the mean of the individual predictions. This averaged prediction $\hat{\Theta}_{I_V} = \frac{1}{8} \sum_{k=1}^{8} \text{CNN}(I_V^k)$ is then used for further analysis to generate virtual 3D microstructures, from which various geometrical descriptors (see Section~\ref{sec:descriptors}) are subsequently computed.

\smallskip

To evaluate accuracy, we used the median absolute error (MdAE) and the median relative absolute error (MdRAE) of the ground truth $y = (y_1, \ldots, y_B)\in\R^B$ and the prediction $\hat{y} = (\hat{y}_1, \ldots, \hat{y}_B)\in\R^B$ for some $B \in \N$, mathematically defined as

\begin{equation*}
    \text{MdAE}(y, \hat{y}) = \text{median}( |y_1 - \hat{y}_1|, \ldots, |y_B - \hat{y}_B|)
\end{equation*}
and 
\begin{equation*}
    \text{MdRAE}(y, \hat{y}) = \text{median}\Bigl( \frac{|y_1 - \hat{y}_1|}{y_1}, \ldots, \frac{|y_B - \hat{y}_B|}{y_B}\Bigr).
\end{equation*}

Using the median instead of the mean, the errors are resistant to outliers, ensuring that few extreme values do not disproportionately influence the results. This approach provides a more reliable assessment of the overall model performance, as it better reflects the typical error. 

\smallskip

Table \ref{tab:descriptors:val} summarizes these errors for each descriptor, computed on the whole validation set $V$.

\begin{table}[H]
    \centering
    \begin{tabular}{llll}
    \toprule
     Descriptor & MdAE & MdRAE\\ 
     \midrule
     $\varepsilon_{\text{CGO}}$ & 0.022 & 7.0 \% \\
     $\varepsilon_{\text{Ni}}$ & 0.018 & 5.8 \% \\
     $\varepsilon_{\text{P}}$ & 0.019 & 6.1 \% \\
     $\mu_{\text{CGO}}$ & 0.024 \si{\micro\meter} & 5.0 \% \\
     $\mu_{\text{Ni}}$ & 0.027 \si{\micro\meter} & 5.1 \% \\
     $\mu_{\text{P}}$ & 0.023 \si{\micro\meter} & 4.7 \% \\
     $\tau_{\text{CGO}}$ & 0.011 & 1.0 \% \\
     $\tau_{\text{Ni}}$ & 0.015 & 1.3 \% \\
     $\tau_{\text{P}}$ & 0.010 & 0.9 \% \\
     $S_{\text{CGO}}$ & 0.104 \si{\per\micro\meter} & 4.4 \% \\
     $S_{\text{Ni}}$ & 0.142 \si{\per\micro\meter} & 5.9 \% \\
     $S_{\text{P}}$ & 0.103 \si{\per\micro\meter} & 4.4 \% \\
     $\rho$ & 0.286 \si{\micro\meter^{-2}} & 8.5 \% \\
     \bottomrule
    \end{tabular}
    \caption{MdAE and MdRAE with respect to the geometrical descriptors described in Table~\ref{tab:descriptors} of the ground-truth and predicted structures.}
    \label{tab:descriptors:val}
\end{table}

\begin{figure}[H]
    \centering
    \begin{subfigure}[b]{\linewidth}
        \centering
        \includegraphics[width=0.8\linewidth]{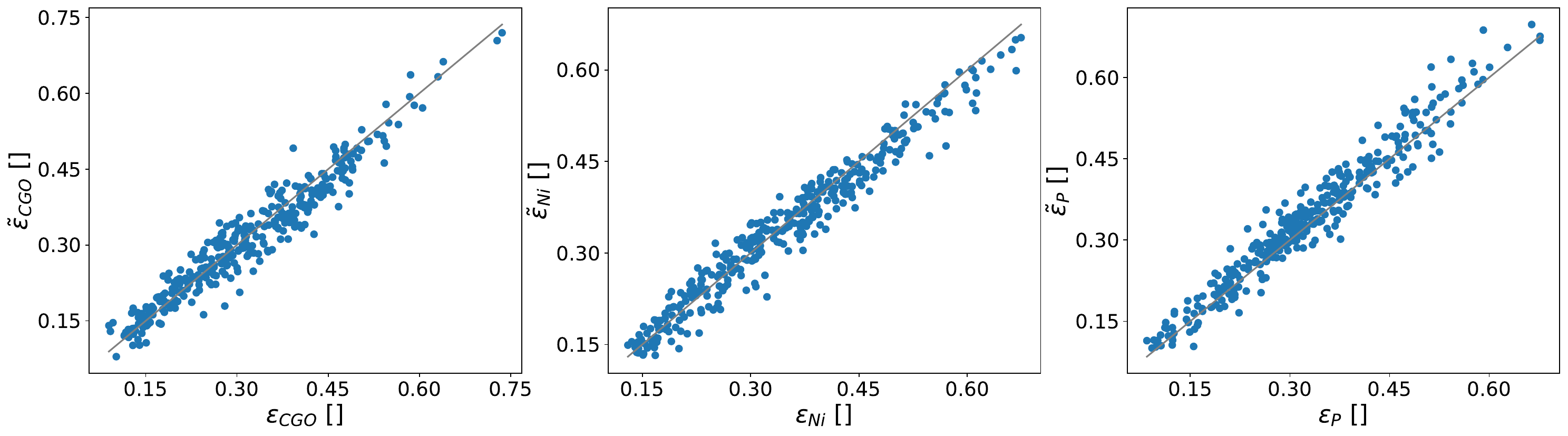}
        \label{fig:err_vol}
    \end{subfigure}
    \begin{subfigure}[b]{\linewidth}
        \centering
        \includegraphics[width=0.8\linewidth]{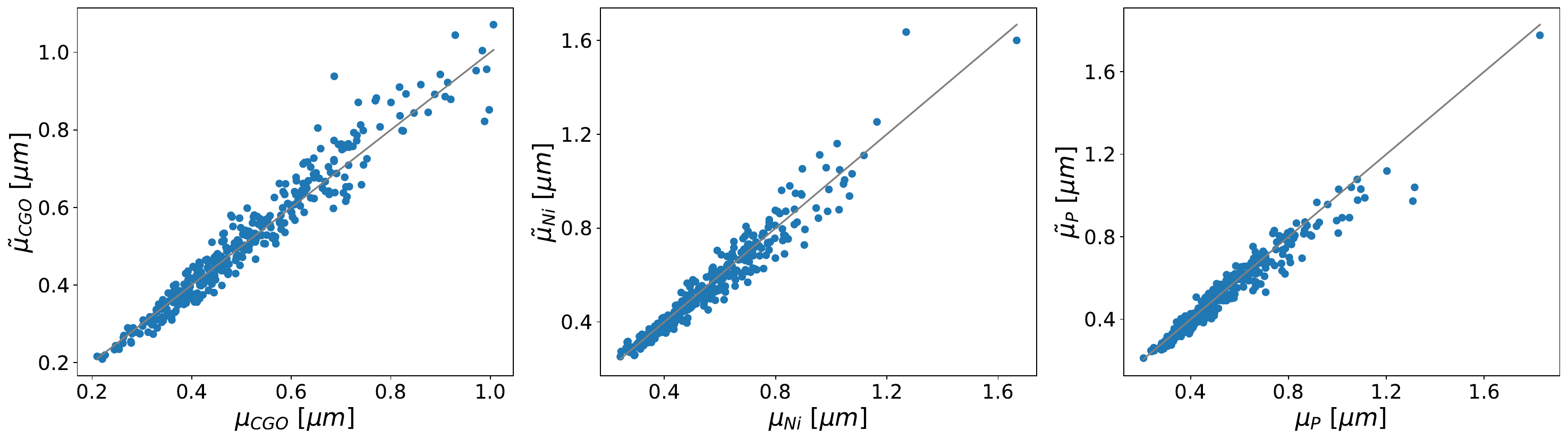}
        \label{fig:err_mcl}
    \end{subfigure}

    \begin{subfigure}[b]{\linewidth}
        \centering
        \includegraphics[width=0.8\linewidth]{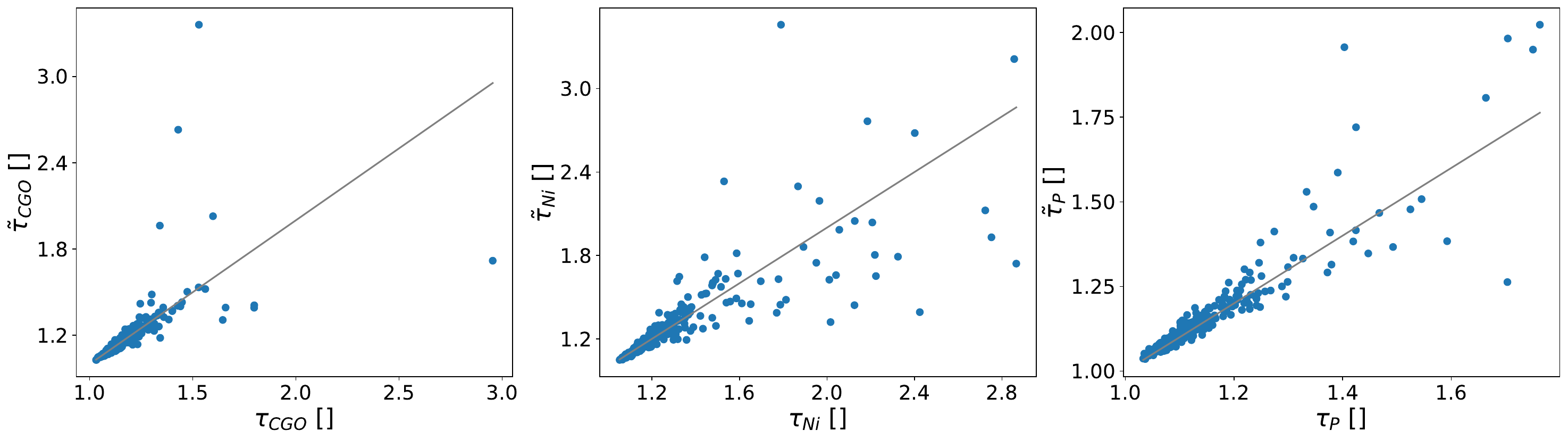}
        \label{fig:err_tort}
    \end{subfigure}
    \begin{subfigure}[b]{\linewidth}
        \centering
        \includegraphics[width=0.8\linewidth]{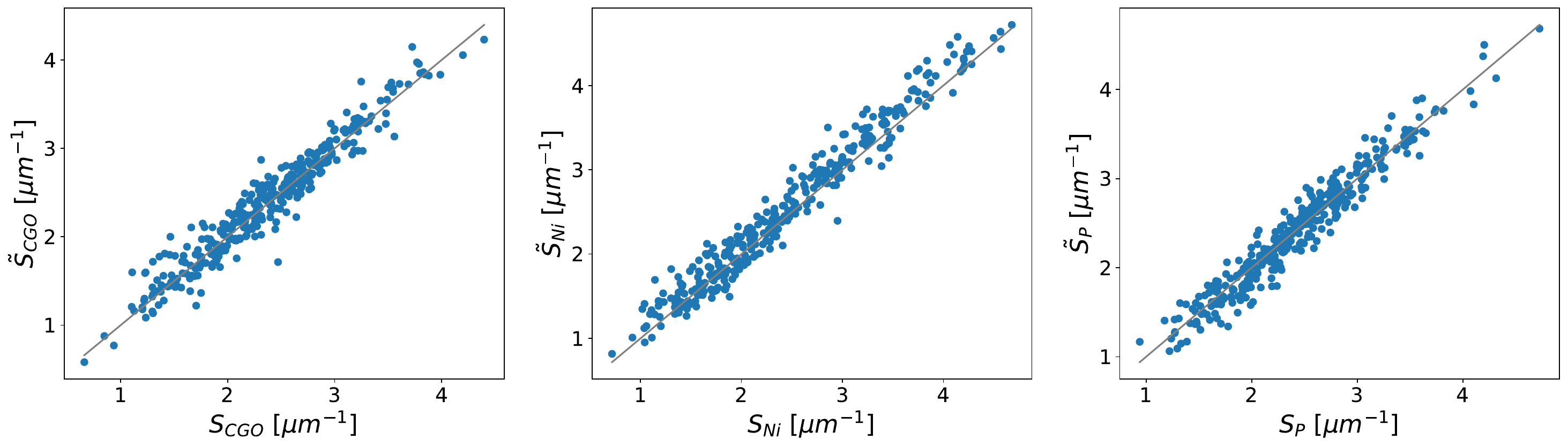}
        \label{fig:err_ss}
    \end{subfigure}
        
    \begin{subfigure}[b]{0.48\linewidth}
        \centering
        \includegraphics[width=0.57\linewidth]{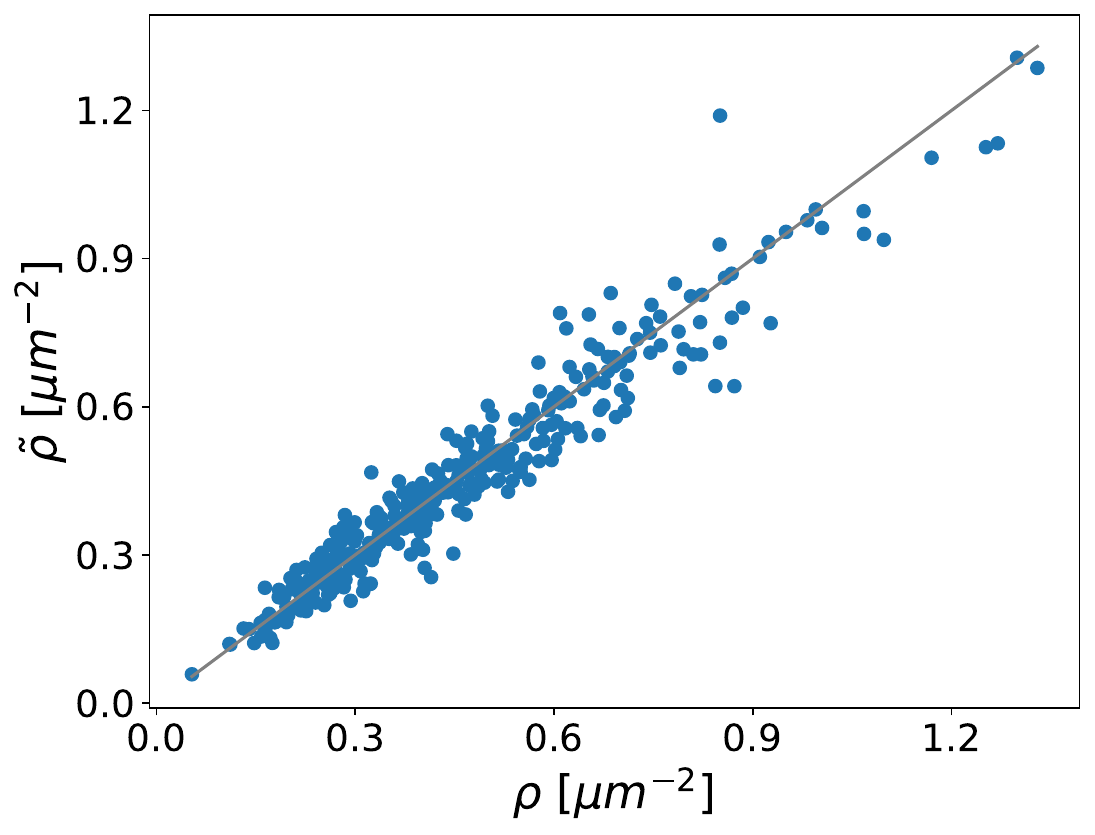}
        \label{fig:err_stpb}
    \end{subfigure}

    \caption{\textbf{Geometrical descriptor analysis of original and generated microstructures.} Comparison of  geometrical descriptors, computed from the original segmented 3D image and the generated structure, respectively: volume fraction (first row), mean chord length (second row), geodesic tortuosity (third row), specific surface area (fourth row), and specific triple phase boundary length (bottom row), where $\varepsilon, \mu, \tau, S, \rho$ are the descriptors computed on the original structure and $\tilde{\varepsilon}, \tilde{\mu}, \tilde{\tau}, \tilde{S}, \tilde{\rho}$ are the descriptors computed on the structure that has been predicted by the CNN. The first four descriptors are  computed for the CGO phase (left), the nickel phase (middle), and the pore space (right).}
    \label{fig:combined_descriptors}
\end{figure}

Overall, the geometrical descriptors match very well, the MdRAE is always below 9\%. Among them, the mean geodesic tortuosities exhibit the best performance, with an MdRAE consistently below 1.3\%. However, as shown in Figure~\ref{fig:combined_descriptors}, we observe that there are some outliers, especially when there is a larger mean geodesic tortuosity in the original 3D microstructure, the prediction is more likely leading to a larger error. For the other four geometrical descriptors, there are no major outliers.

\subsection{Performance on real microstructure data}
\begin{figure}[H]
    \centering
    \includegraphics[width=0.45\linewidth]{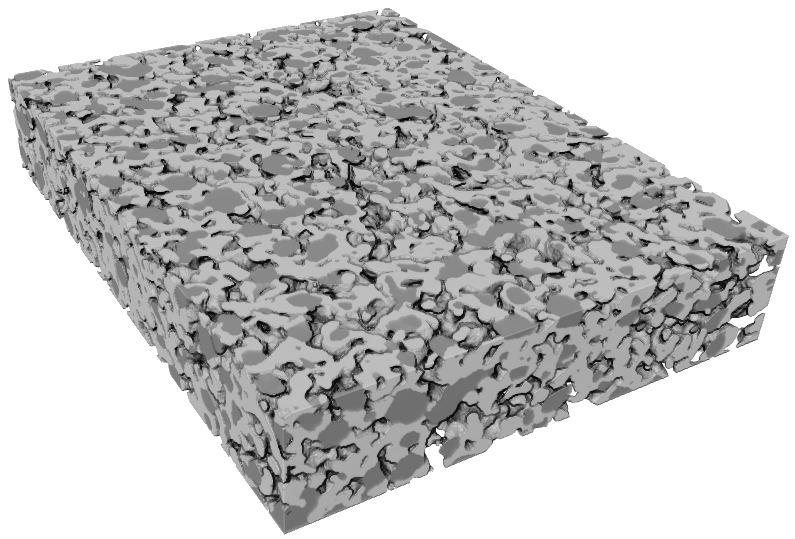}
    \caption{\textbf{3D rendering of tomographic image data.} SOFC anode of size $\SI{37}{\micro\meter} \times \SI{28.15}{\micro\meter} \times \SI{6.4}{\micro\meter}$, where the transparent, light gray and dark gray colors correspond to pore, CGO and nickel phase, respectively.}
    \label{fig:orignal_sample}
\end{figure}

To assess the generalizability of our model to real-world data, we evaluated its performance on a real 3D microstructure. For this, we used a segmented 3D image (see Figure~\ref{fig:orignal_sample}), which has been obtained via 3D FIB-SEM (sample A in~\cite{weber_descriptors}),  and simulated the corresponding SEM image. This image was then processed in the same way as the images in the validation set $V$, i.e., eight random cutouts of the image were extracted and then processed by the CNN. The final prediction of the model parameters was obtained by averaging the eight predictions. Using these predicted parameters, we generated a 3D microstructure and computed geometrical descriptors for both the original segmented 3D image and the generated structure.

The geometrical descriptors computed from the original and the predicted structures match very well,  see Table~\ref{tab:descriptors:test}. 
The only descriptor where the prediction deviates more than 9\% from the ground truth is the specific surface area of the pore space $S_{\text{P}}$ with a deviation of approximately 18\%.

\begin{table}[H]
    \centering
    \begin{tabular}{llll}
    \toprule
     Descriptor & Real & Predicted & Relative error\\ 
     \midrule
     $\varepsilon_{\text{CGO}}$ [ ]& 0.42 & 0.40 & 3.4\,\%\\
     $\varepsilon_{\text{Ni}}$ [ ]& 0.22 & 0.21 & 4.8\,\%\\
     $\varepsilon_{\text{P}}$ [ ]& 0.36 & 0.39 & 6.9\,\%\\
     $S_{\text{CGO}}$ [\si{\per\micro\meter}] & 2.22  & 2.26 & 1.6\,\%\\
     $S_{\text{Ni}}$ [\si{\per\micro\meter}] & 1.08  & 1.08 & 0.2\,\%\\
     $S_{\text{P}}$ [\si{\per\micro\meter}] & 1.89  & 2.24 & 18.1\,\%\\
     $\mu_{\text{CGO}}$ [\si{\micro\meter}] & 0.65 & 0.68 & 3.7\,\% \\
     $\mu_{\text{Ni}}$ [\si{\micro\meter}] & 0.69 & 0.75 & 8.7\,\% \\
     $\mu_{\text{P}}$ [\si{\micro\meter}] & 0.68 & 0.66 & 3.2\,\% \\
     $\tau_{\text{CGO}}$ [ ] & 1.12 & 1.09 & 2.2\,\%\\
     $\tau_{\text{Ni}}$ [ ] & 1.61 & 1.48 & 8.2\,\%\\
     $\tau_{\text{P}}$ [ ] & 1.11 & 1.10 & 0.7\,\% \\
     $\rho$ [\si{\micro\meter^{-2}}] & 0.21 & 0.22 & 5.7\,\% \\
     \bottomrule
    \end{tabular}
    \caption{Comparison of selected geometrical descriptors of the real and the predicted structure.}
    \label{tab:descriptors:test}
\end{table}

Note that some deviation in the geometrical descriptors is expected since even a perfect prediction of the parameters of the stochastic 3D microstructure model by the CNNs would lead to deviations caused by the following two reasons. First, multiple realizations of a stochastic 3D geometry model differ slightly in terms of their geometrical descriptors due to the inherent randomness. Second, the real 3D microstructure of the SOFC anode is not guaranteed to be adequately modeled via the excursion-set model described in Section~\ref{sec:model_description}.

\section{Conclusion}
\label{sec:conclusion}

We implemented a CNN-based approach to predict the parameters of a stochastic 3D model, enabling the prediction of 3D microstructures from 2D SEM images. To train this network, we first used the stochastic model to generate 2,000 synthetic microstructures, which we then processed via an SEM simulation tool to obtain realistic 2D SEM images. These parameter-image pairs served as our training data. Using the trained network, we predicted the model parameters from SEM images and generated the corresponding virtual 3D microstructures. Finally, we quantitatively validated our approach by comparing  geometrical descriptors of these predicted structures with the original ones.

\smallskip

This method provides several advantages over traditional approaches. By relying solely on 2D SEM images, we avoid the expensive acquisition of 3D image data. Additionally, the presented framework avoids the phase-based segmentation of 2D images, further  simplifying the workflow and reducing potential sources of error. 

\smallskip

In future work, we plan to incorporate secondary electron (SE) SEM images as an additional input channel for our neural network. This could further improve its performance by providing complementary information in conjunction with existing 
backscattered electron (BSE) images.
Furthermore, this approach is adaptable for different materials with potentially different stochastic microstructure models. Moreover, when other imaging techniques like X-ray CT are the state of the art for a material, this approach allows to substitute the SEM simulator with a corresponding simulator, for example a simulator for X-ray CT images.

\section*{Funding}
The authors would like to thank the  German Federal Ministry of Research, Technology and Space (BMFTR) for financial support within the project ``Datengetriebene Struktur-Eigenschafts-Analyse mittels stochastischer und numerischer Simulationsmethoden - Workflow zur Strukturoptimierung von SOFC-Brennstoffzellen (DASEA-4-SOFC)'' (grant number 05M22VUA). The funder played no role in study design, data collection, analysis and interpretation of data, or the writing of this manuscript.

\section*{Data availability}
The datasets generated and analysed during the current study are not publicly available, since they are part of an ongoing research project, but are available from the corresponding author on reasonable request.

\section*{Code availability}
All formulations and algorithms necessary to reproduce the results of this study are described in Sections 2, 3 and 4.
The underlying code for this study is not publicly available but may be made available to qualified researchers on reasonable request from the corresponding author.

\section*{Author Contributions}
L.F.S. contributed across all aspects of the study, including SEM simulation, CNN training, analysis, and manuscript preparation. S.W. and B.P. contributed to stochastic modeling. L.F. assisted with the implementation and training of the CNN. V.S. supported the work through funding acquisition, project administration and supervision. B.P. supervised the study. All authors discussed the results and contributed to writing the manuscript.

\section*{Competing interests}
The authors declare no competing interests.

\bibliographystyle{unsrt}
\bibliography{bibliography}

\end{document}